\documentclass[pdftex,twocolumn,epjc3]{svjour3}          % twocolumn

\RequirePackage[T1]{fontenc}

\smartqed  % flush right qed marks, e.g. at end of proof

\RequirePackage{graphicx}
\RequirePackage{mathptmx}      % use Times fonts if available on your TeX system
\RequirePackage{flushend}
\RequirePackage[numbers,sort&compress]{natbib}
\RequirePackage[colorlinks,citecolor=blue,urlcolor=blue,linkcolor=blue]{hyperref}
\RequirePackage{multirow}
\RequirePackage{morefloats}
\RequirePackage{amssymb,amsmath}
\RequirePackage{cases}
\journalname{Eur. Phys. J. C}

\begin{document}

\title{Maximum mass of an anisotropic compact object admitting the modified Chaplygin equation of state in Buchdahl-I metric
}


\author{D. Bhattacharjee\thanksref{e1,addr1}, 
	P. K. Chattopadhyay\thanksref{e2,addr1} 
	}

%\thankstext[$\star$]{t1}{Thanks to the title}
\thankstext{e1}{e-mail: debadriwork@gmail.com}
\thankstext{e2}{e-mail: pkc$_{-}$76@rediffmail.com}

\institute{IUCAA Centre for Astronomy Research and Development (ICARD), Department of Physics, Cooch Behar Panchanan Barma University, Vivekananda Street, District: Coochbehar,  Pin: 736101, West Bengal, India.\label{addr1} }
\date{Received: date / Accepted: date}

\maketitle

\abstract{In this article, a new class of exact solutions for anisotropic compact objects is presented. Admitting the modified Chaplygin equation of state $p=H\rho-\frac{K}{\rho^{n}}$, where $H$, $K$ and $n$ are constants with $0<n\leq1$, and employing the Buchdahl-I metric within the framework of the general relativity stellar model is obtained. Recent observations on pulsars and GW events reveal that the observed maximum mass of compact stars detected so far is approximately $2.59^{+0.08}_{-0.09}~M_{\odot}$. Since massive stars cannot be supported by a soft equation of state, a constraint of the equation of state must hold. The choice of a suitable equation of state for the interior matter of compact objects may predict useful information compatible with recent observations. TOV equations have been solved using the modified Chaplygin equation of state to find the maximum mass in this model. In particular, the theory can achieve $3.72~M_{\odot}$, when $H=1.0$, $K=10^{-7}$ and $n=1$. The model is suitable for describing the mass of pulsars PSR J2215+5135 and PSR J0952-0607 and the mass $2.59^{+0.08}_{-0.09}~M_{\odot}$ of the companion star in the GW 190814 event. The $3.72~M_{\odot}$ is hardly achievable theoretically in general relativity considering fast rotation effects too. To check the physical viability of this model, we have opted for the stability analysis and energy conditions. We have found that our model satisfies all the necessary criteria to be a physically realistic model.}

\section{Introduction}
\label{sec1}
Stellar evolution is one of the most fascinating phenomena in astrophysics. Observational results of the stellar bodies have provided the pathway to make new theories to study stellar properties. With the development of general relativity, Einstein opened new vistas to test various theories in the grandest of scales. In general, compact objects are the end state of stellar evolution. Therefore, the study of compact objects provides theoretical insights into highly dense matter configurations. In the last decade, progress in the area of theoretical modeling has made it smoother to transition from a theoretical perspective to analytical solutions. In the recent past, a large number of data from pulsars and GW events have been collected and analysed by many investigators. Their analysis predicted accurate estimations of many physical parameters of such astrophysical objects. The accurate estimation of the mass of a compact object with the help of the Shapiro delay \cite{Cromartie} yielded a mass of $(2.14)^{+0.10}_{-0.09}~M_{\odot}$ for the compact object millisecond pulsar PSR J0740+6620. Since then, the topic of the maximum mass of compact objects has also been discussed. Apart from that, it is also argued that a handful of compact objects may exist that may achieve greater masses than the mass of PSR J0740+6620 \cite{Cromartie} in the group of interacting systems. The gravitational wave event denoted as GW 190814 has shown that the companion star has a mass of $2.5-2.67~M_{\odot}$ with 90\% confidence \cite{Abbott}. However, it is not clear to astrophysicists whether this component is a very massive compact star or  a light black hole. Hence, if the first possibility is correct, then it is obviously necessary to include a new concept in the theory to increase the range of maximum mass to accommodate such high mass values. Such formalism would be important for the study of the internal composition of dense matter above saturation density.

In this context, anisotropy in pressure plays a crucial role in the maximum mass and radius of compact objects, as the maximum mass and stability both increase with increasing anisotropy. Following the work of Ruderman \cite{Ruderman} and Canuto \cite{Canuto}, it is observed that anisotropy in the high density regime $(\rho>10^{15}gm/cm^{3})$ may appear locally and play a very important role in describing the properties of the interior matter of compact objects. The origin of anisotropy in highly dense compact objects can be explained with the natural examples such as fermionic fields, electromagnetic fields in Neutron stars \cite{Sawyer}, Pion condensation \cite{Sawyer1}, Superfluidity \cite{Carter} etc.. In the review article\cite{Mardan}, the origin of the local anisotropy was investigated in detail. Additionally, it was established that the possible source of local anisotropy may be viscosity \cite{Herrera}.

To generalise the models of compact stars, Bowers and Liang \cite{Bowers} imposed the idea of pressure anisotropy and explored the dependence of stellar properties such as compactness, and the mass-radius of the star, which exhibit a high redshift. Heintzmann and Hillebrandt \cite{Heintzmann} studied the relativistic properties of anisotropic neutron stars and found that in the context of arbitrarily large anisotropic factors, there is, in principle, neither mass nor a redshift limit. A considerable number of works \cite{Maurya1}-\cite{Hernandez} have been devoted to studying the spherically symmetric anisotropic stellar configuration in static equilibrium after the pioneering work of Carter and Langlois \cite{Carter}. In addition, Kalam et al. \cite{Kalam1} showed the dependence of central density on the value of the anisotropy parameter. Anisotropic spherically symmetric modeling provides a more generalised way to employ the equation of state (henceforth EoS). If the EoS is known, we can also distinctively use the Tolman-Oppenheimer-Volkoff equation \cite{Tolman}-\cite{Oppenheimer} to determine the internal structure of the stellar configuration as well as their maximum possible mass and radius.

It is well established that our present universe is passing through an accelerated phase and that the ordinary matter and fields of standard cosmology are not sufficient to sustain the theories of an expanding universe. Significant modifications regarding the matter distributions in Einstein gravity theories are necessary to withstand the present-day observational cosmological results. In this backdrop, a new notion for matter distribution has come up that must exert negative pressure. Interestingly, exotic matter contains negative pressure, and until now, the exact exotic matter EoS has not been identified. There are many EoSs that are used to describe exotic matter, and the Chaplygin gas EoS is one such prime example. The Chaplygin gas EoS \cite{Chaplygin}-\cite{Karman} is represented as $p=-\frac{A}{\rho}$, where $A>0$ and $p$ and $\rho$ are known as pressure and energy density, respectively. This EoS has been generalised by Bento et al. \cite{Bento} in the form, $p=-\frac{A}{\rho^{n}}$, where $n$ is a free parameter characterised by a range $0<n\leq1$. Consequently, $n=1$ generates the original Chaplygin EoS. A modified form of the Chaplygin gas EoS (henceforth MCG) has been considered by Liu and Li \cite{Jun} in the context of cosmology given as, $p=H\rho-\frac{K}{\rho^{n}}$. The MCG EoS is a more generalised form taking three free parameters into consideration, and it also covers the whole aspect of the original Chaplygin gas EoS. Inside the very massive compact objects, it may be possible that exotic matter may exist, which may be characterised by MCG EoS. H.B. Benaoum \cite{Benaoum} studied the accelerated universe under the framework of the FRW metric incorporating MCG EoS. Gorini et al. \cite{Gorini} studied the solutions of Tolman-Oppenheimer-Volkoff equations in static spherically symmetric space-time, including both the phantom and non-phantom cases. Thakur et al. \cite{Bcp} explored MCG as a viable choice for dark energy and obtained the numerical constraints of the free parameters. Bhar et al. \cite{Bhar} constructed an anisotropic compact star model where the interior matter sector is characterised by MCG and the obtained mass and radius results from the model are compared with the observational results to a high degree of accuracy. To find the exact solutions of EFE, the choice of metric potential is very much important. For the spherically symmetric distribution of perfect fluid in static equilibrium, Delgaty and Lake \cite{Delgaty} tabulated a list of metric ansatzes to evaluate the exact or closed solutions of EFE. Such solutions are important to predict viable physical features of compact objects. The Buchdahl-I metric \cite{Buchdahl} is very useful for studying the properties of compact objects in this context. Durgapal and Banerji \cite{Durgapal} rederived the Buchdahl-I metric ansatz \cite{Buchdahl} for the analytical modeling of relativistic fluid spheres in spherically symmetric space-time. Maurya et al. \cite{Maurya5} studied the anisotropic compact star in the Buchdahl metric ansatz with a simplified notion that the EoS can be approximated as a linear function of energy density. In another study, Maurya et al. \cite{Maurya6} used the Buchdahl metric ansatz to study the hydrostatic equilibrium conditions for stellar structures within the framework of modified $f(R,T)$ gravity theory.

Recent observations on pulsars and GW events along with the latest accurate measurement of the maximum mass of a compact object using Shapiro delay yields $2.14^{+0.1}_{-0.09}~M_{\odot}$ \cite{Cromartie} for millisecond pulsar MSP J0740+6620. Detection of GW event GW190814 reveals that one component of the binary system may have a mass of approximately $2.59^{+0.08}_{-0.09}~M_{\odot}$, which has puzzled the astrophysicist community. Therefore, keeping in mind the stand of measurements, compact stars may not be all made of self-bound hadronic matter, even considering the anisotropic effect, which increases the maximum mass limit fairly above the $2~M_{\odot}$ figure. Therefore, necessary theory is essential to incorporate such a mass limit. Therefore, it would be very important for the composition of interior matter or simply the microphysics of the dense matter higher than the saturation density. Recently, the fastest and heaviest pulsar PSR J0952-0607 detected in the disk of the Milky Way galaxy of mass $2.35~M_{\odot}$ may contain strange quark matter in its composition, as observational evidence supports it \cite{Carvalho}. From the theory of the normal neutron star model in GR, the prediction of such a high value of maximum mass is hardly obtainable even considering fast rotation effects.

The above observational evidence motivated us to revisit a class of self-bound stellar models taken into consideration in the last couple of decades to allow feasible stellar sequences to act in accordance with such a high maximum mass. The solutions are presented and will be used to make comparisons with recently published data for compact objects. The basic aim of this study is to construct a suitable stellar model to explain the properties of newly observed high-mass compact objects and their physical features, such as maximum mass, radius, energy density $(\rho)$, radial $(p_{r})$ and transverse $(p_{t})$ pressures and pressure anisotropy $(\Delta)$. Using the model, the radii of many pulsars and secondary objects of GW events may be predicted.

The paper is organised as follows: In section~(\ref{sec2}), a spherically symmetric line element is considered, and we solve the Einstein field equations incorporating both the MCG EoS \cite{Benaoum} and Buchdahl metric ansatz \cite{Buchdahl} for the $g_{rr}$ component. In this section, we have determined the corresponding metric potential $(\nu)$, energy density $(\rho)$, radial $(p_{r})$ and tangential $(p_{t})$ pressures and also the anisotropic factor $(\Delta)$. In section~(\ref{sec3}), we have matched the interior solutions with the exterior vacuum geometry to compute the constants present in the ansatz. Section~(\ref{sec4}) provides the necessary limits on the values of free MCG parameters for a physically realistic model. In section~(\ref{sec5}), we graphically show the relationship between the maximum mass and radius for this model. Here, we have also predicted the radius of some recently observed compact objects and compared them with the available values estimated from observations. Section~(\ref{sec6}) deals with the graphical representation of basic properties of a stellar configuration. Section~(\ref{sec7}) is used to present the viability of the model through the causality criteria. The energy conditions and their radial variations are depicted through graphical representations in section~(\ref{sec8}). In section~(\ref{sec9}), we analyse the stability of the model through the well-established methods of the generalised TOV equation, the cracking condition proposed by Herrera, the value of the adiabatic index and stability against small radial oscillation through the Lagrangian perturbation procedure, and it is found that the model obeys all the necessary stability conditions. Finally, we conclude by discussing the main findings in this model in section~(\ref{sec10}).
                   
\section{Einstein field equations and their solutions with the modified Chaplygin EoS}
\label{sec2}
The spherically symmetric space-time in static equilibrium is characterised by the line element given below: 
\begin{equation}
	ds^2=-e^{2\nu(r)}dt^2+e^{2\lambda(r)}dr^2+r^2(d\theta^2+sin^2\theta d\phi^2). \label{eq1}
\end{equation}
The Einstein's field equations (henceforth EFE) connecting the matter and geometry are expressed as:
\begin{equation}
	R_{ij}-\frac{1}{2}g_{ij}R=8\pi T_{ij}, \label{eq2}
\end{equation}
In relativistic units, $G=c=1$ and $T_{ij}$ is the stress-energy tensor for the static matter distribution. The most general anisotropic form of $T_{ij}$ is given by 
\begin{equation}
	T_{ij}=diag(-\rho,p_{r},p_{t},p_{t}). \label{eq3}
\end{equation}
Using Eqs.~(\ref{eq1}) and (\ref{eq3}) in Eq.~(\ref{eq2}), the EFEs are given in the following form
\begin{eqnarray}
	\frac{2e^{-2\lambda}\lambda'}{r}+\frac{(1-e^{-2\lambda})}{r^2}=8\pi\rho, \label{eq4} \\
	\frac{2e^{-2\lambda}\nu'}{r}-\frac{(1-e^{-2\lambda})}{r^2}=8\pi p_{r}, \label{eq5} \\
	e^{-2\lambda}(\nu''+\nu'^2-\lambda'\nu'+\frac{\nu'}{r}-\frac{\lambda'}{r})=8\pi p_{t},\label{eq6}
\end{eqnarray}
where overhead prime $(')$ denotes derivative w.r.t. $r$. In this formulation, we consider the form of the Buchdahl-I metric \cite{Buchdahl} as given below \cite{Delgaty}:
\begin{equation}
	e^{2\lambda(r)}=\frac{2(1+\chi r^2)}{2-\chi r^2}, \label{eq7}
\end{equation}
where $\chi$ is a constant whose dimension is $Km^{-2}$. Using Eq.~(\ref{eq7}) in Eq.~(\ref{eq4}), we express the energy density $(\rho)$ in the form: 
\begin{equation}
	\rho=\frac{3\chi(3+\chi r^2)}{16\pi(1+\chi r^2)^2}. \label{eq8}
\end{equation}
The modified Chaplygin equation of state (henceforth EoS) is given in the following form \cite{Benaoum}:
\begin{equation}
	p_{r}=H\rho-\frac{K}{\rho^n}, \label{eq9}
\end{equation}
where, $H$ and $K$ are positive constants. Throughout this model formalism, we consider $n=1$, therefore, Eq.~(\ref{eq9}) takes the form:
\begin{equation}
	p_{r}=H\rho-\frac{K}{\rho}. \label{eq10}
\end{equation}
Here, $H$ is dimensionless, whereas the dimension of $K$ is $Km^{-4}$. Using Eqs.~(\ref{eq7}), (\ref{eq8}) and Eq.~(\ref{eq10}) in Eq.~(\ref{eq5}), we have solved Eq.~(\ref{eq5}) to obtain the value of metric potential $\nu$ in the following form: 
\vspace{0.25cm}
\begin{eqnarray}
\nu=\frac{1}{60\chi^2}\Big((-15\chi^2(5H+3)+6912\pi^2 K)log(2-\chi r^2)\nonumber \\+30H\chi^2log(\chi r^2+1)  +128\pi^2K(5\chi r^2(\chi r^2+4)\nonumber \\+16log(\chi r^2+3))\Big), \label{eq11}
\end{eqnarray}
where we have considered the integration constant to be zero. Using Eq.~(\ref{eq8}) in Eq.~(\ref{eq10}), we obtain the radial pressure $p_{r}$:
\begin{equation}
	p_{r}=\frac{3\chi H(\chi r^2+3)}{16\pi(1+\chi r^2)^2}-\frac{16\pi K(1+\chi r^2)^2}{3\chi(\chi r^2+3)}. \label{eq12}
\end{equation}
Similarly using Eqs.~(\ref{eq7}) and (\ref{eq11}), we can obtain the expression for tangential pressure $p_{t}$ from Eq.~(\ref{eq6}) in the following form:
\vspace{0.5cm}
\begin{eqnarray}
	p_{t} = -\frac{1}{14400\pi}\Big(A+B+\frac{1638400K^2\pi^4r^4}{\chi} \nonumber\\	+1638400K^2\pi^4r^6+C+D+E+F+G+I\Big), \label{eq13}
\end{eqnarray}
\\
where, \\ \\
$
A=-\frac{12800K\pi^2(3\chi^2(9H+8)+640K\pi^2)}{\chi^3}, \nonumber \\
B=\frac{38400K\pi^2(\chi^2(1-3H)+384K\pi^2)r^2}{\chi^2},\nonumber \\
C=\frac{6(25\chi^2(H+1)-2304K\pi^2)(5\chi^2(5H+3)-2304K\pi^2)}{\chi^3(\chi r^2-2)}, \nonumber \\
D=\frac{2700H\chi(H-3)}{(1+\chi r^2)^3}, \nonumber \\
E=\frac{450\chi(3+H(2H+11))}{(1+\chi r^2)^2},\nonumber \\
F=-\frac{75\chi(H+1)(23H+3)}{(1+\chi r^2)}, \nonumber \\
G=\frac{61400K\pi^2(15\chi^2-512K\pi^2)}{\chi^3(\chi r^2+3)^2}, \nonumber \\
I=\frac{1024K\pi^2(80896K\pi^2-135\chi^2)}{\chi^3(\chi r^2+3)}.\nonumber \\
 $
 \\
\\
The anisotropy factor $(\Delta)$ is defined as the difference between $p_{t}$ and $p_{r}$, i.e.
\begin{equation}
	\Delta=p_{t}-p_{r}. \label{eq14}
\end{equation}
At the center of the star, $\Delta=0$, i.e., $p_{t}=p_{r}$. \\ The total gravitational mass contained within the sphere of radius $R$ is obtained as:
\begin{equation}
	m(r)=4\pi\int_{0}^{R} \rho r^2 dr. \label{eq15}
\end{equation}
\section{Boundary Condition} 
\label{sec3}
We match the interior space-time with the exterior Schwarzs\\-child space-time at the surface of the compact object to corelate and evaluate the constants in the metric elements. The exterior Schwarzschild metric is given as:
\begin{equation}
	ds^2=-\Big(1-\frac{2M}{r}\Big)dt^2+\frac{1}{(1-\frac{2M}{r})}dr^2+r^2(d\theta^2+sin^2\theta d\phi^2). \label{eq16}
\end{equation}
At the surface $(r=R)$ of the compact object, the continuity of the two metric potentials yields,  
\begin{equation}
	e^{-2\lambda}=1-2u, \label{eq17}
\end{equation}
and \begin{equation}
	e^{2\nu}= 1-2u. \label{eq18}
\end{equation}
Here, $u=\frac{M}{R}$ is the compactness of the star. Again, the boundary of the star is defined as the surface where radial pressure drops to zero, i.e.,
\begin{equation}
	p_{r}(R)=0. \label{eq19}
\end{equation}
Now, using Eqs.~(\ref{eq16})-(\ref{eq18}), we determine the constants $\chi$ and $K$ in the form:
\begin{equation}
	\chi=\frac{4u}{R^2(3-4u)}, \label{eq20} 
\end{equation}
\begin{equation}
	K=\frac{9\chi^{2}H(3+\chi R^{2})^{2}}{256\pi^{2}(1+\chi R^{2})^{4}}. \label{eq20a}
\end{equation}
\section{Bounds on the modified Chaplygin EoS parameters $H$ and $K$}\label{sec4} For a physically realistic model, the energy density and pressure must be positive and finite at the center ($r=0$). Using Eq.~(\ref{eq8}), the central density can be written as:
\begin{equation}
	\rho_{0}=\frac{9\chi}{16\pi}=\frac{9u}{4\pi R^2(3-4u)}, \label{eq21}
\end{equation}
Using Eq.~(\ref{eq12}), the expression for central pressure is: 
\begin{equation}
	p_{r}(0)=\frac{9\chi H}{16\pi}-\frac{16\pi K}{9\chi}. \label{eq22}
\end{equation}
From Eq.~(\ref{eq21}), the positivity of the energy density is only ensured for $\chi>0$. Within this notion, for the positive central pressure, it is noted that in this model, there exist some bounds on parameters $H$ and $K$, which obey the following equality:
\begin{equation}
	\frac{H}{K}>\frac{256\pi^{2}}{81\chi^2}, \label{eq23}
\end{equation}
i.e., the ratio $(\frac{H}{K})$ depends on the compactness $u$ and radius $R$ of the star. Thus, the values of $H$ and $K$ could not be chosen arbitrarily. We have chosen the values of $H$ and $K$ obeying Eq.~(\ref{eq23}).
\section{Mass-radius relation from the TOV equation} \label{sec5} 
\begin{figure}[ht!]
	%	\centering
	\includegraphics[width=8cm]{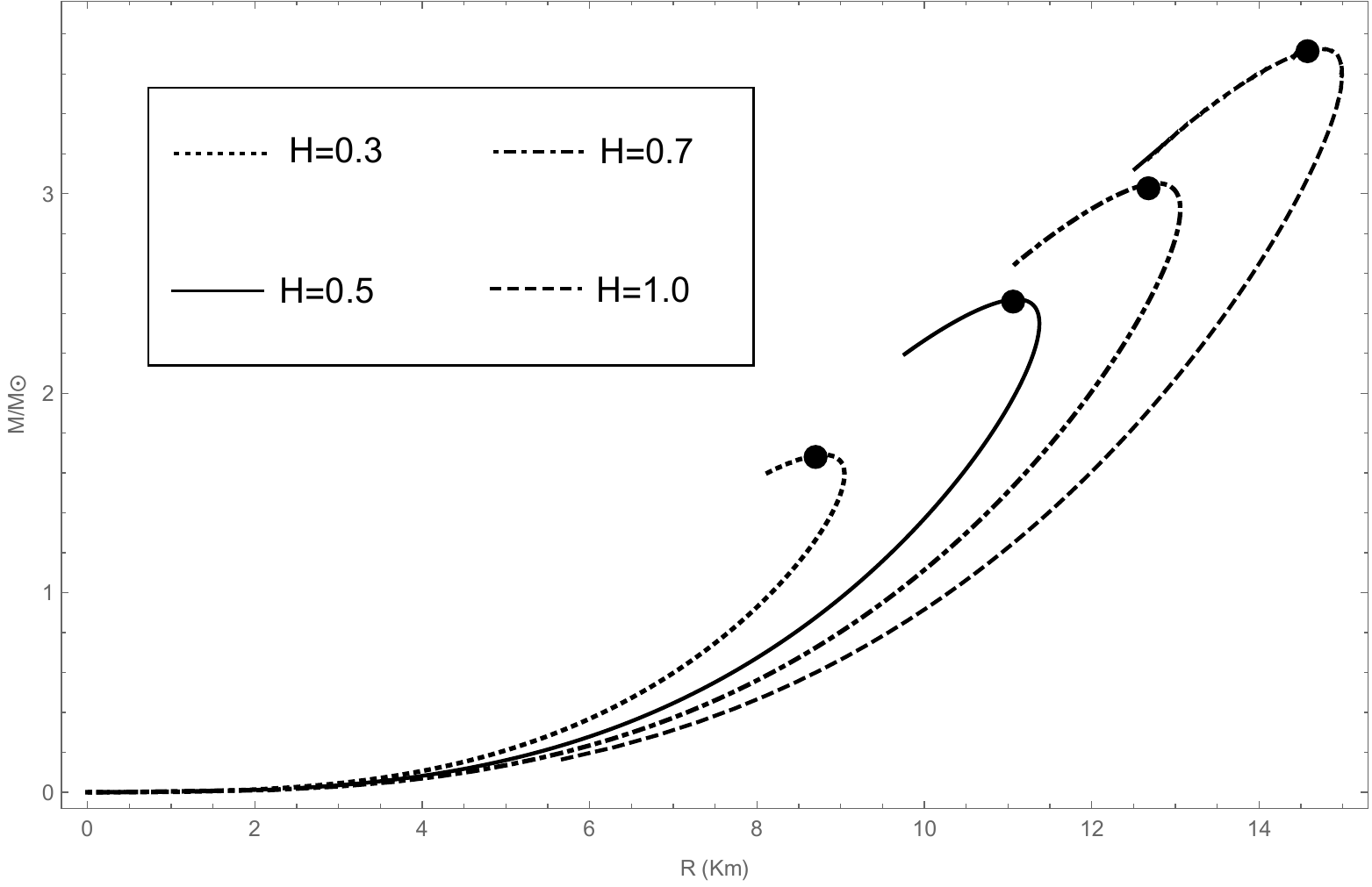}
	\caption{Mass radius plot for $K=10^{-7}$ and different values of $H$.} 
	\label{fig1}
\end{figure}

Following the work mentioned in Ref. \cite{Bcp1} and using the condition of Eq.~(\ref{eq23}), we have solved the TOV \cite{Tolman,Oppenheimer} equations to determine the mass-radius relation of compact objects in this model and are plotted in Fig.~(\ref{fig1}) for the parametric choice of $H$ and $K$. We have obtained a wide range of maximum mass $(M_{max})$ from $1.18~M_{\odot}-3.72~M_{\odot}$ and their corresponding radii from $7.179~Km-14.72~Km$ for $n=1$, $K=10^{-7}$ and $H=0.2-1.0$. The range of parameter $H$  is taken from Ref. \cite{Bcp1} and $K=10^{-7}$. From Tab.~(\ref{tab1}) it is noted that the maximum mass and radius both increase with increasing $H$. Apart from this range $(H=0.2-1.0)$, it is not possible to solve the TOV equations in the present context.
\begin{table}[ht!]
	\centering 
	\caption{MCG EoS parameters and maximum mass-radius (K fixed at $10^{-7}$)}
	\label{tab1}
	\begin{tabular*}{\columnwidth}{@{\extracolsep{\fill}}ccc@{}}
		\hline
		H  & Max Mass ($M/M_{\odot}$) & Radius(km)  \\
		\hline
		0.20 & 1.18 & 7.179 \\
		
		0.30 & 1.69 & 8.79 \\
		
		0.40 & 2.11 & 10.05 \\
		
		0.50 & 2.47 & 11.10  \\
		
		0.56 & 2.66 & 11.66 \\
		
		0.60 & 2.78 & 11.99  \\
		
		0.70 & 3.05 & 12.78 \\
		
		0.80 & 3.29 & 13.48  \\
		
		0.90 & 3.52 & 14.16  \\
		
		1.00 & 3.72 & 14.72  \\ 
		\hline
	\end{tabular*}	
\end{table}
It is evident that the employment of the MCG equation of state may increase the maximum mass range, which means that compact objects can contain more mass against their gravitational collapse in the presence of MCG. In the following table, we present a list of pulsars and secondary objects of GW events whose mass and radius may be predicted as well as compared from our formalism. From Tab.~(\ref{tab2}), it is evident that one may predict the radius of a large number of pulsars and secondary objects of GW events by adjusting the value of constants $H$ and $K$. Thus, the model is suitable for determining the radii of a wider class of compact objects.
\begin{table*}[ht!]
	\centering 
	\caption{Predicted radius (Km) of some recently observed pulsars and companion objects of GW events $(K=10^{-7})$ from our model. }
	\label{tab2}
	\begin{tabular}{ccccccc}
		\hline
		Compact Object  & Measured Mass  & Measured Radius & \multicolumn{4}{c}{Predicted Radius from model (Km)} \\ \cline{4-7} 
		& ($M/M_{\odot}$) & (Km) &H=0.50 & H=0.56 & H=0.70 & H=1.0 \\
		\hline
		GW 190814 \cite{Abbott} & $2.59^{+0.08}_{-0.09}$ & -- & -- & 11.90 & 12.83 & 13.90\\
		
		PSR J0952-0607 \cite{Carvalho} & 2.35 & -- & 11.37 & 11.83 & 12.54 & 13.53\\
		
		PSR J0030+0451 \cite{Miller} & 1.44 & 13.02 & 10.15 & 10.38 & 10.85 & 11.58\\
				
		PSR J0740+6620 \cite{Riley} & 2.072 & 12.39 & 11.18 & 11.50 & 12.11 & 12.99\\
		
		GW 170817 \cite{Bauswein} & 1.4 & -- & 10.04 & 10.30 & 10.75 & 11.50\\
		
		PSR J1614-2230 \cite{Demorest} & 1.97 & 11-13 &  11.05 & 11.36 & 11.95 & 12.80 \\
		
		PSR J2215+5135 \cite{Linares} & 2.27 & -- &  11.35  & 11.75 & 12.42 & 13.38\\
		
		4U 1608-52 \cite{Guver} & 1.74 & 9.3$\pm$1.0 & 10.07 & 10.97 & 11.50 & 12.32\\
		\hline
	\end{tabular}	
\end{table*}
\section{Physical application of the model}
\label{sec6} 
In this section, we have analysed the behavior of the basic compact object parameters such as energy density $\rho$, radial pressure $p_{r}$, tangential pressure $p_{t}$ and anisotropy factor $\Delta$ with radius $r~(Km)$. For physical application, we have considered PSR J0740+6620 with a  mass of 2.072$M_{\odot}$ and a radius of 12.39 Km \cite{Riley} and a corresponding compactness of u=0.246. For physical analysis, we consider $H=0.3$, and using Eq.~(\ref{eq20a}), we have computed the value of $K$ for some known compact objects and are tabulated in Tab.~(\ref{tab3}). It is noted that the value of $K$ depends on $H$ and the stellar mass and radius. The corresponding central density, surface density and central pressure are also tabulated in Tab.~(\ref{tab3}). 
 
\begin{table*}[ht!]
	\centering 
	\caption{Evaluation of physical parameters of few known compact objects}
	\label{tab3}
	\begin{tabular}{cccccccc}
		\hline
		Compact Object & Mass ($M_{\odot}$) & Radius (Km) & H & K & Central density $(\rho_{c}) $ & Surface density $(\rho_{s})$ & Central pressure $(p_{c})$ \\
		&&&&&$(g/cm^{3})$ & $(g/cm^{3})$ & $(dyn/cm^{2})$ \\
		\hline
		PSR J0030+0451 & 1.44 & 13.02 & & 0.115$\times10^{-7}$ & 0.396$\times10^{15}$ & 0.265$\times10^{15}$ & 0.594$\times10^{35}$ \\
		PSR J0740+6620 & 2.072 & 12.39 & 0.3 & 0.269$\times10^{-7}$ & 0.771$\times10^{15}$ & 0.404$\times10^{15}$ & 1.51$\times10^{35}$  \\
		4U 1608-52 & 1.74 & 9.3 & & 0.991$\times10^{-7}$ & 1.62$\times10^{15}$ & 0.775$\times10^{15}$ & 3.39$\times10^{35}$ \\
		\hline
	\end{tabular}	
\end{table*}  
\begin{figure}[ht!]
		%\centering
		\includegraphics[width=8cm]{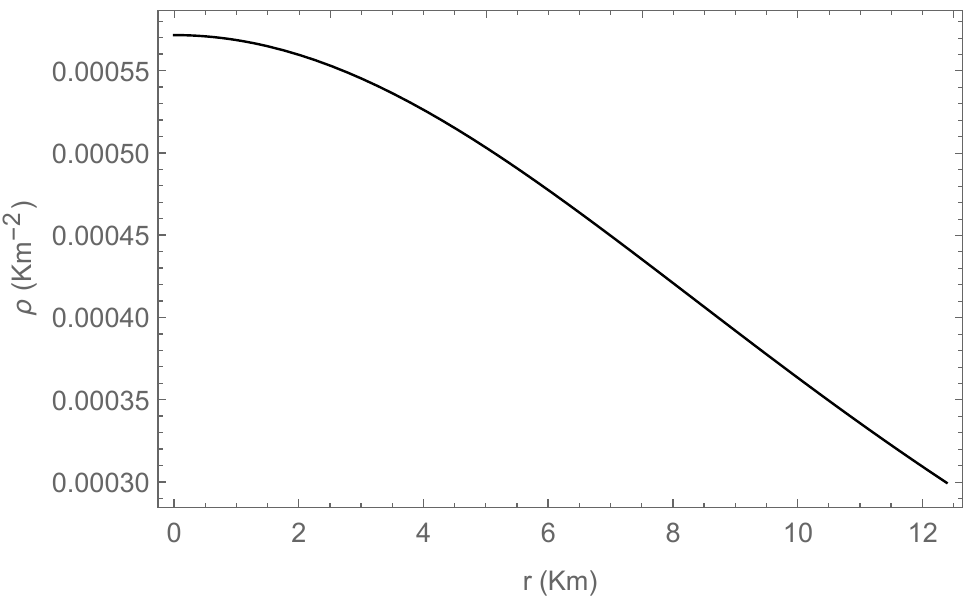}
		\caption{Radial variation of energy density $(\rho)$ for $H=0.3$ and $K=0.269\times10^{-7}$.}
		\label{fig2}
	\end{figure}
	\begin{figure}[ht!]
		%\centering
		\includegraphics[width=8cm]{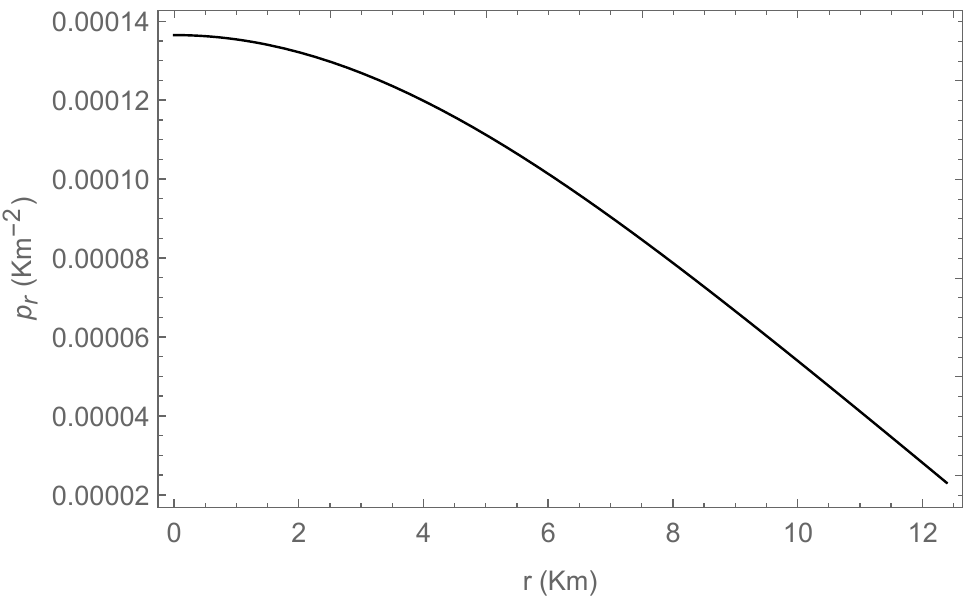}
		\caption{Radial variation of radial pressure $(p_{r})$ for $H=0.3$ and $K=0.269\times10^{-7}$.}
		\label{fig3}
\end{figure}
\begin{figure}[ht!]
		%\centering
		\includegraphics[width=8cm]{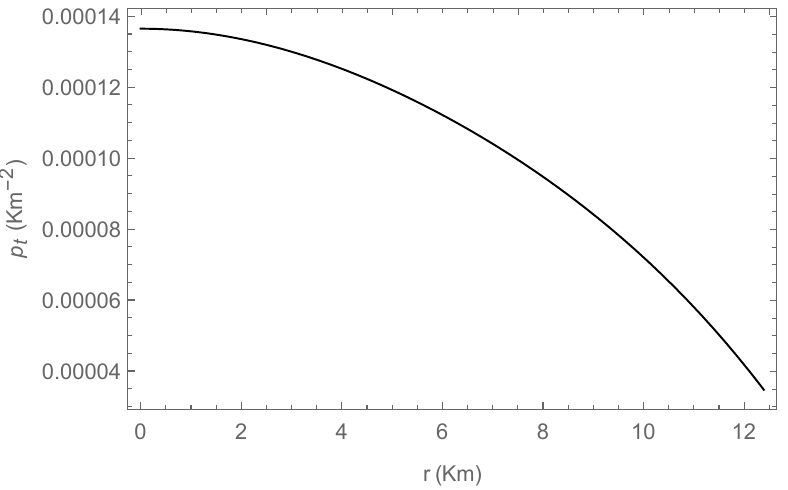}
		\caption{Radial variation of tangential pressure $(p_{t})$ for $H=0.3$ and $K=0.269\times10^{-7}$.}
		\label{fig4}
\end{figure}
\begin{figure}[ht!]
		%\centering
		\includegraphics[width=8cm]{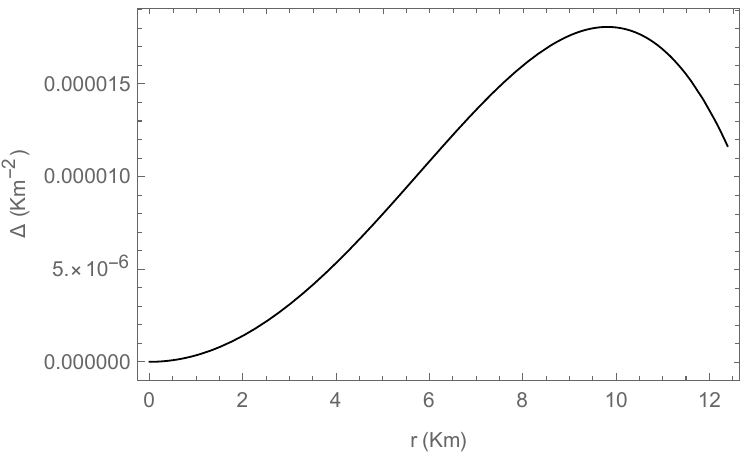}
		\caption{Radial variation of anisotropy $(\Delta)$ for $H=0.3$ and $K=0.269\times10^{-7}$.}
		\label{fig5}
\end{figure}

From Figs.~(\ref{fig2})-(\ref{fig4}) we note that the energy density, radial pressure and tangential pressure decrease from the center to the surface which is a viable condition for a stable compact stellar configuration.
\subsection{Causality Condition} \label{sec7} 
\begin{figure}[h!]
		%\centering
		\includegraphics[width=8cm]{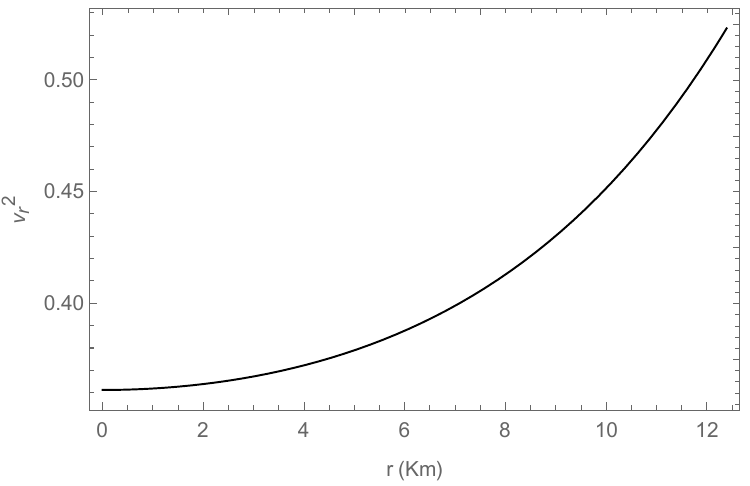}
		\caption{Variation of $(v_{r}^{2})$ vs $r$ for $H=0.3$ and $K=0.269\times10^{-7}$.}
		\label{fig6}
\end{figure}

\begin{figure}[h!]
		%\centering
		\includegraphics[width=8cm]{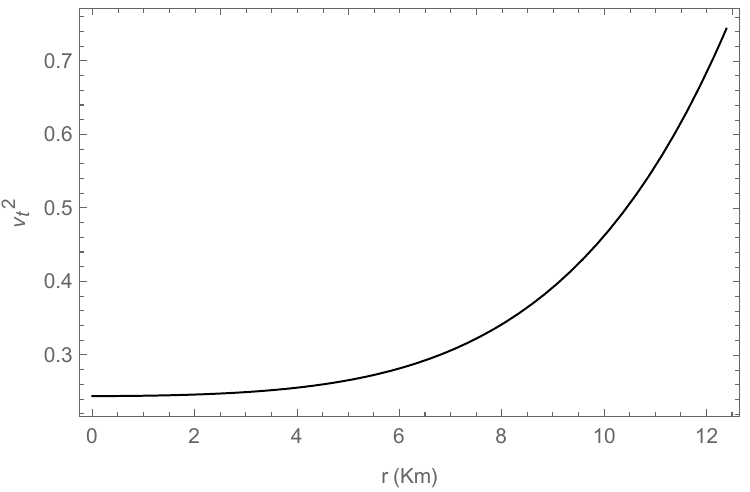}
		\caption{Variation of $(v_{t}^{2})$ vs $r$ for $H=0.3$ and $K=0.269\times10^{-7}$.}
		\label{fig7}
\end{figure}

For a realistic model of an anisotropic compact star, one way of characterising its interior dense matter is through the study of the velocity of sound waves given by $v_{r}^{2}=(\frac{dp_{r}}{d\rho})$ and $v_{t}^{2}=(\frac{dp_{t}}{d\rho})$, where $\rho$ is the energy density including the rest mass energy of the constituent particles, and $p_{r}$ and $p_{t}$ represent radial and tangential pressures, respectively. Here, we use the system of units as $h=c=1$. The causality condition on sound velocities implies an absolute upper bound as $v_{r}^{2}\leq1$ and $v_{t}^{2}\leq1$. On the other hand, the thermodynamic stability ensures that $v_{r}^{2}>0$ and $v_{t}^{2}>0$. Therefore, within the stellar composition, the conditions $0<v_{r}^{2}\leq1$ and $0<v_{t}^{2}\leq1$ should hold simultaneously. Due to the complexity of the expressions of sound velocities, we have shown the variations of $v_{r}^{2}$ and $v_{t}^{2}$ graphically in Figs.~(\ref{fig6}) and (\ref{fig7}), respectively. 
It is evident from Figs.~(\ref{fig6}) and (\ref{fig7}) that the causality conditions are well obeyed in this model. 
\subsection{Energy Conditions}\label{sec8}In gravitational theory, the energy conditions are imposed on matter distributions to obtain a physically viable energy momentum tensor. Qualitatively, these conditions are a way to seek the nature of matter distribution without the need for explicit specifications about the internal matter content. Hence, it is possible to obtain the physical features of extreme events, {\it viz.}, gravitational collapse or existence of geometrical singularity, etc., without the knowledge of energy density or pressure. In essence, the study of energy conditions is an algebraic problem \cite{Kolassis}, more specifically the eigenvalue problem of the energy momentum tensor. In 4-dimensional space-time, the study of energy conditions leads to the roots of a 4-degree polynomial, which is complicated due to the presence of analytical solutions of eigenvalues. Even though the general solution is difficult to obtain, a physically realistic fluid distribution should follow the null, weak, strong and dominant energy conditions \cite{Kolassis}-\cite{Wald} simultaneously within the stellar boundary. Here, we have studied the energy conditions \cite{Brassel,Brassel1} for the present stellar configuration and found that they are well satisfied.

In the case of a physically viable model of a compact star, the necessary energy conditions \cite{Brassel}, \cite{Brassel1} must be fulfilled at all internal points as well as at the surface of compact stars. For the present model, we have checked the energy conditions within the parameter space used here and found that they are in good agreement with the prescribed conditions. We have shown the energy energy conditions through graphical representation. The study includes the verification of the following energy conditions: 
\begin{enumerate}
	\item Null energy condition (NEC): $\rho+p_{r}\geq0, \rho+p_{t}\geq0 $. \\
	\item Weak energy condition (WEC): $\rho\geq0, \rho+p_{r}\geq0, \rho+p_{t}\geq0 $. \\
	\item Strong energy condition (SEC): $\rho+p_{r}\geq0, \rho+p_{t}\geq0, \rho+p_{r}+2p_{t}\geq0$. \\
	\item Dominant Energy Condition (DEC): $\rho\geq0, \rho-p_{r}\geq0, \rho-p_{t}\geq0 $.
\end{enumerate}  
	\begin{figure}[h!]
	%	\centering
		\includegraphics[width=8cm]{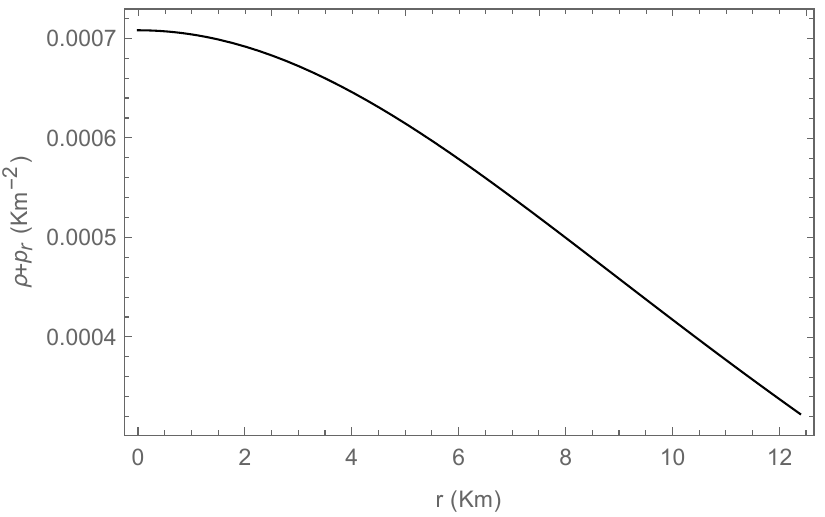}
		\caption{Radial variation of $(\rho+p_{r})$ for $H=0.3$ and $K=0.269\times10^{-7}$.} 
		\label{fig8}
		\end{figure}
	\begin{figure}[h!]
	%	\centering
		\includegraphics[width=8cm]{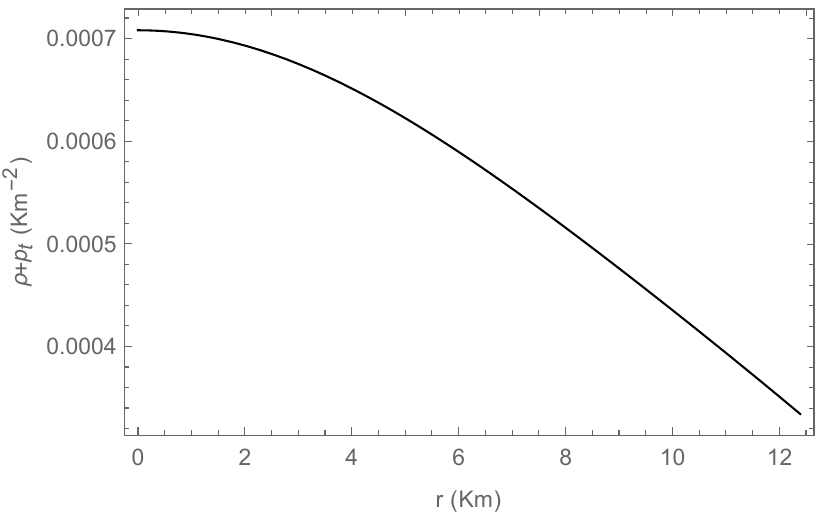}
		\caption{Radial variation of $(\rho+p_{t})$ for $H=0.3$ and $K=0.269\times10^{-7}$.} 
		\label{fig9}
\end{figure}
\begin{figure}[h!]
		%\centering
		\includegraphics[width=8cm]{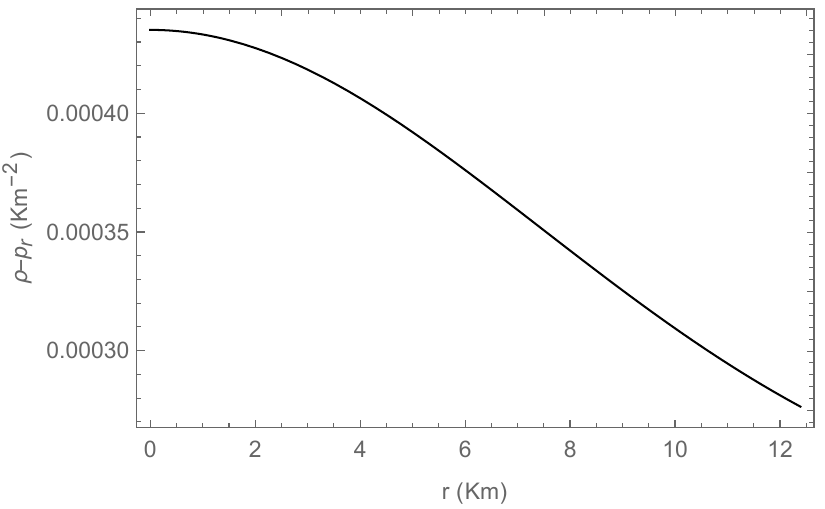}
		\caption{Radial variation of $(\rho-p_{r})$ for $H=0.3$ and $K=0.269\times10^{-7}$.} 
		\label{fig11}
\end{figure}
\begin{figure}[h!]
	%	\centering
		\includegraphics[width=8cm]{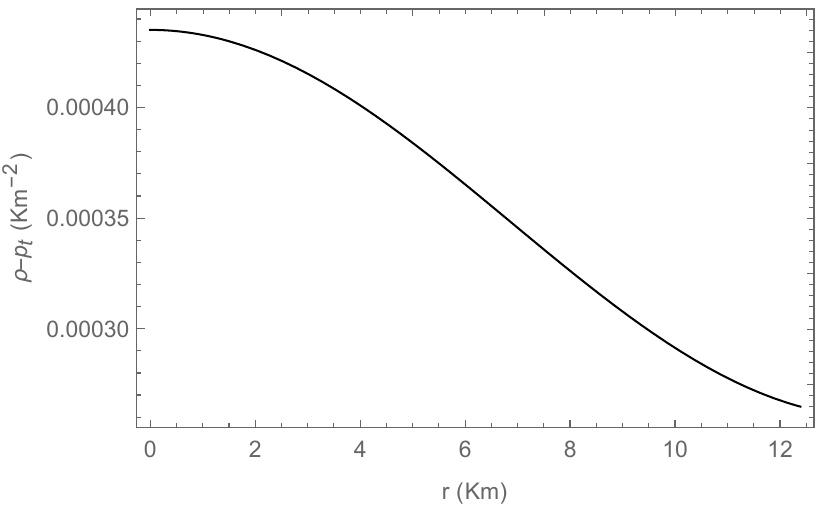}
		\caption{Radial variation of $(\rho-p_{t}$) for $H=0.3$ and $K=0.269\times10^{-7}$.} 
		\label{fig12}
\end{figure}
\begin{figure}[h!]
	%	\centering
	\includegraphics[width=8cm]{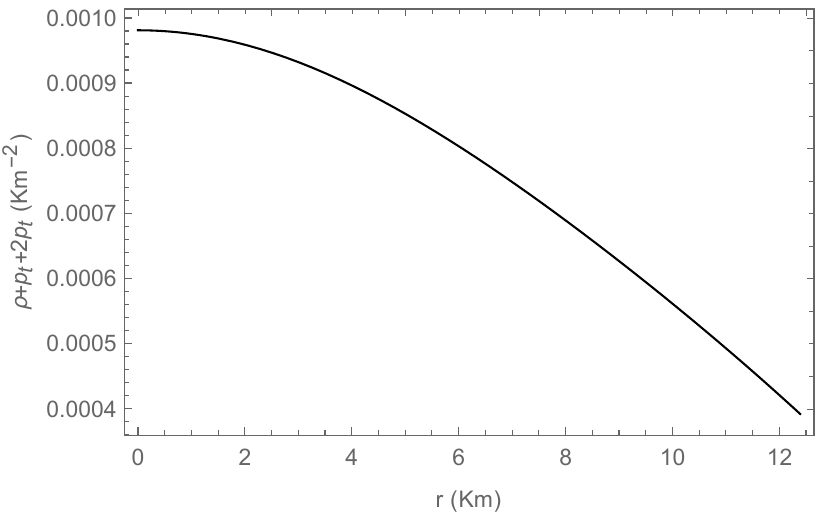}
	\caption{Radial variation of $(\rho+p_{r}+2p_{t})$ for $H=0.3$ and $K=0.269\times10^{-7}$.} 
	\label{fig10}
\end{figure}
The fulfillment of the above energy conditions is shown in Figs.~(\ref{fig2}) and (\ref{fig8}-\ref{fig12}).
\section{Stability Analysis}\label{sec9} The stability of this model is explored on the basis of the following methods: \\
(i) Generalized TOV equation, \\
(ii) Cracking condition proposed by Herrera, \\
(iii) Variation of the adiabatic index and \\
(iv) Lagrangian oscillation.
\subsection{Generalised TOV equation} It is important to study the stability of a model under the influence of different forces. For an anisotropic compact object, the stability analysis is based on the following force components- (i) the gravitational force ($F_{g}$), (ii) the hydrostatic force ($F_{h}$) and (iii) the anisotropic force ($F_{a}$). The model should be in equilibrium under the combined influence of these forces. In this context, we have studied the stability using the generalised Tolman-Oppenheimer-Volkoff (TOV) equation \cite{Tolman,Oppenheimer} of the form given below:
\begin{equation}
	-\frac{M_{G}(r)(\rho+p_{r})}{r^2}e^{\lambda-\nu}-\frac{dp_{r}}{dr}+\frac{2\Delta}{r}=0, \label{eq24}
\end{equation} 
where, $M_{G}$ is referred to as the active gravitational mass derived from the mass formula of Tolman-Whittaker \cite{Gron} given as: 
\begin{equation}
	M_{G}(r)=r^{2}\nu'e^{\nu-\lambda}. \label{eq25}
\end{equation}
Substituting Eq.~(\ref{eq25}) in Eq.~(\ref{eq24}), we obtain
\begin{equation}
	-\nu'(\rho+p_{r})-\frac{dp_{r}}{dr}+\frac{2\Delta}{r}=0. \label{eq26}
\end{equation}  
Here, 
\begin{equation}
	F_{g}=-\nu'(\rho+p_{r}),  \label{eq27} 
\end{equation}
\begin{equation}
	F_{h}=-\frac{dp_{r}}{dr},  \label{eq28}
\end{equation}	
and
\begin{equation}
	F_{a}=\frac{2\Delta}{r}.  \label{eq29}
\end{equation}
Using Eqs.~(\ref{eq8}), (\ref{eq12}) and (\ref{eq13}), we can compute the expressions of Eqs.~(\ref{eq27})-(\ref{eq29}). We have chosen to represent the equilibrium conditions of the model through graphical representation. 
\begin{figure}[h!]
%	\centering
	\includegraphics[width=8cm]{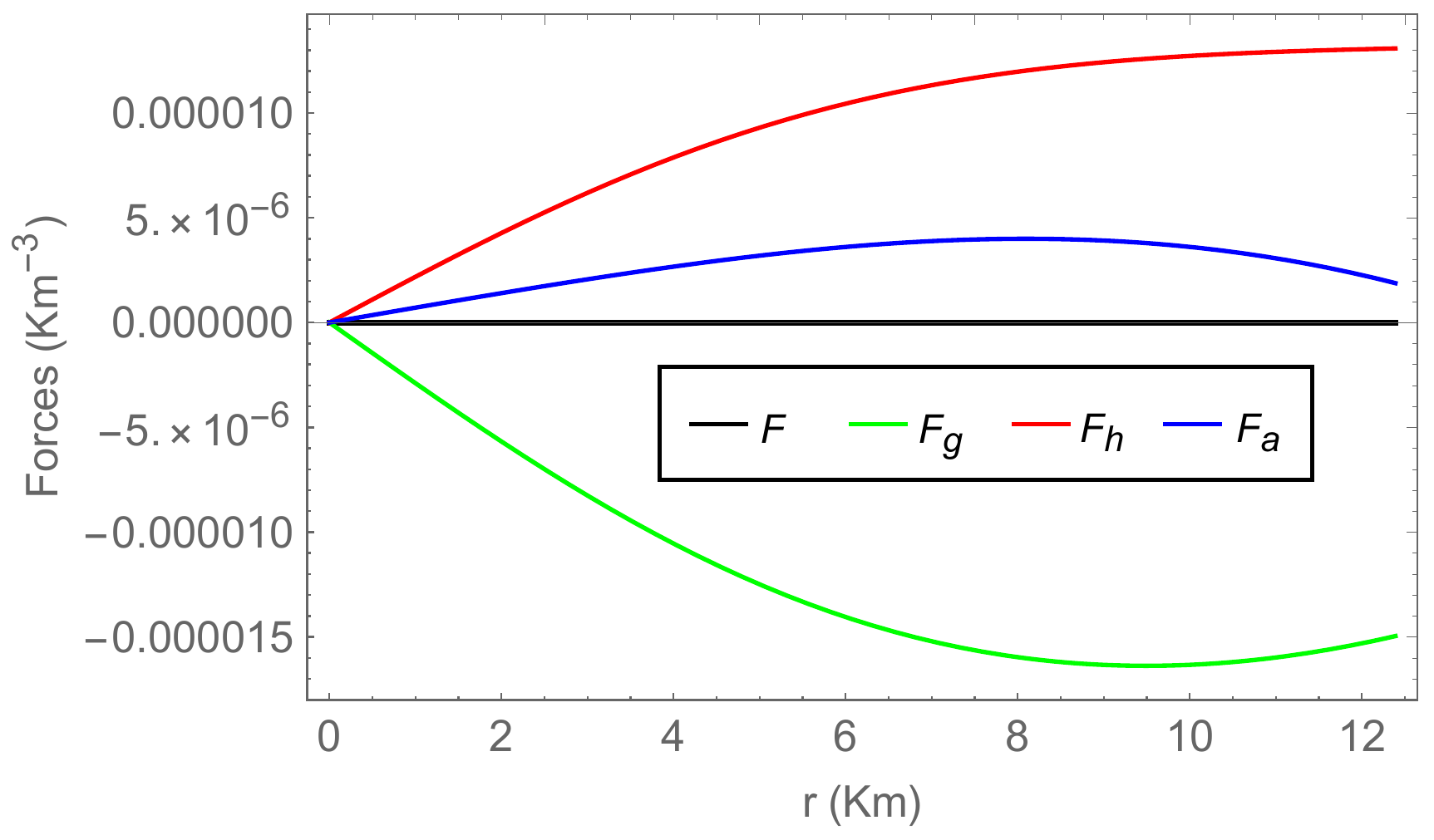}
	\caption{Variation of different forces with radial distance r for $H=0.3$ and $K=0.269\times10^{-7}$.} 
	\label{fig13}
\end{figure}
\subsection{Cracking condition proposed by Herrera} Anisotropic models should be stable under fluctuations in their physical parameters. Herrera \cite{Herrera1} instigated a "cracking" condition to check the stability of such models. On the basis of Herrera's concept Abreu et al. \cite{Abreu} put forward a criterion that determines the stability of an anisotropic stellar model if according to Abreu et al. \cite{Abreu}, the square of radial velocity ($v_{r}^{2}$) and tangential velocity ($v_{t}^{2}$) obey the condition
\begin{equation}
	0\leq|v_{r}^{2}-v_{t}^{2}|\leq 1, \label{eq30}
\end{equation}
thus, it may be termed a stable structure. 
\begin{figure}[h!]
	%\centering
	\includegraphics[width=8cm]{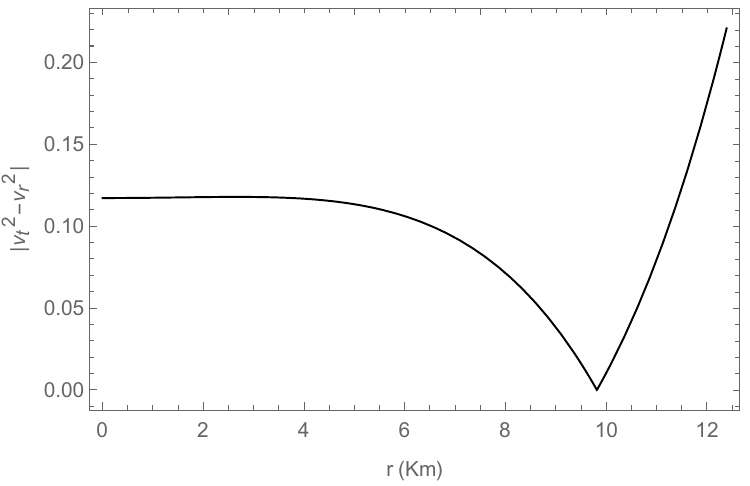}
	\caption{Variation of $|v_{r}^{2}-v_{t}^{2}|$ with radial distance r for $H=0.3$ and $K=0.269\times10^{-7}$.} 
	\label{fig14}
\end{figure}
From Fig.~(\ref{fig14}), it is noted that the Abreu inequality given in Eq.~(\ref{eq30}) is satisfied at all points of PSR J0740+6620. Therefore, the numerical choice of $H$ and $K$ parameter values is viable.  
\subsection{Variation of the adiabatic index} The adiabatic index for the relativistic anisotropic stellar model is expressed as:
\begin{equation}
	\Gamma=\frac{\rho+p_{r}}{p_{r}}\frac{dp_{r}}{d\rho}=\frac{\rho+p_{r}}{p_{r}}v_{r}^{2}. \label{eq31}
\end{equation}
According to the work of Heintzmann and Hillebrandt \cite{Heintzmann}, the condition for the stability of an isotropic stellar model is represented as $\Gamma>\frac{4}{3}$ (Newtonian limit). Furthermore, in the case of an anisotropic star (both $p_{r}$ and $p_{t}$ exist), such a condition is modified by Chan et al. \cite{Chan} and is given as:
\begin{equation}
	\Gamma>\Gamma'_{max}, \label{eq32}
\end{equation}
where, 
\begin{equation}
	\Gamma'_{max}=\frac{4}{3}-\Bigg[\frac{4}{3}\frac{(p_{r}-p_{t})}{|p'_{r}|r}\Bigg]_{max}. \label{eq33}
\end{equation}
\begin{figure}[h!]
	%\centering
	\includegraphics[width=8cm]{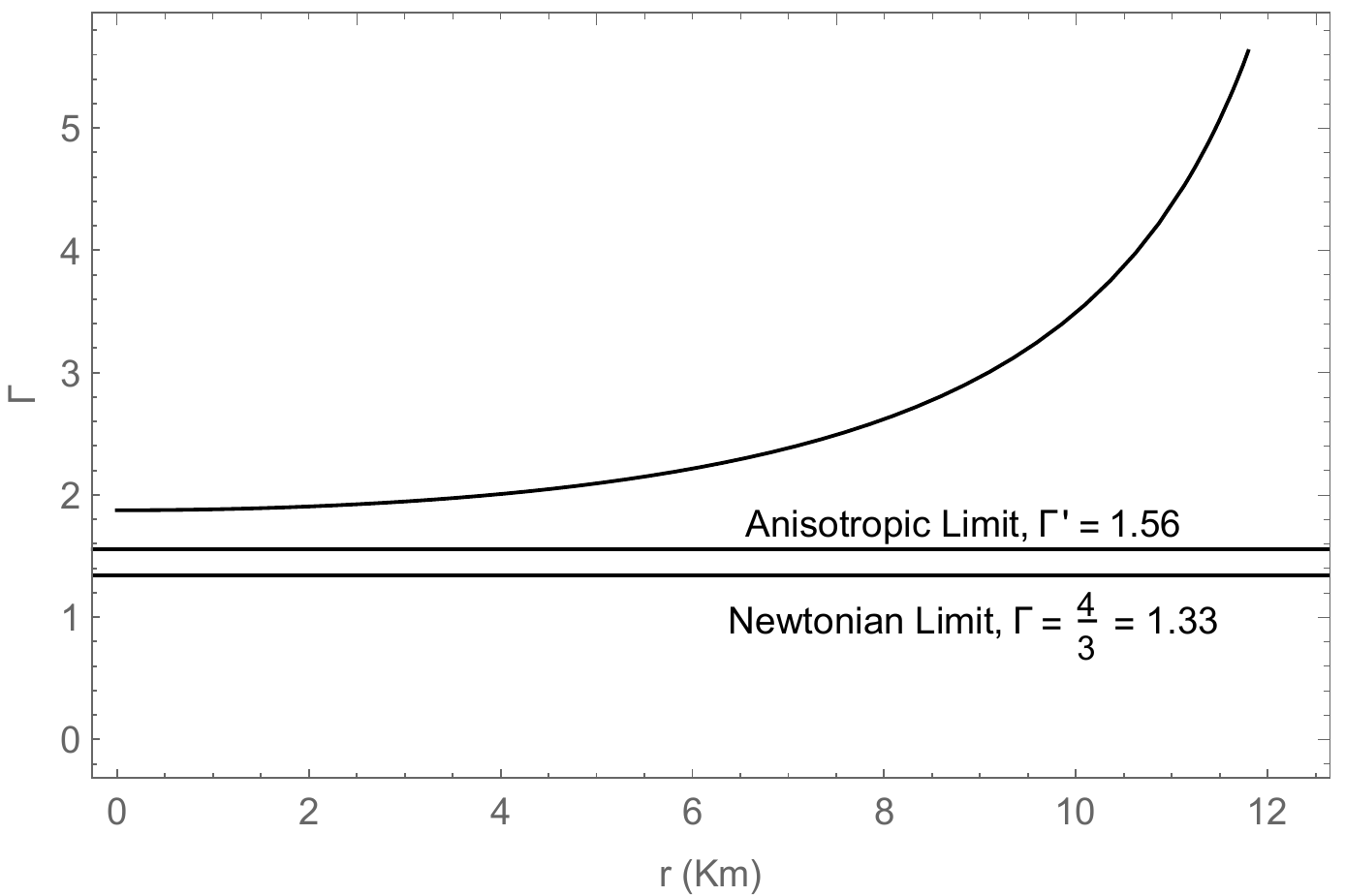}
	\caption{Radial variation of the adiabatic index $(\Gamma)$ for $H=0.3$ and $K=0.269\times10^{-7}$.} 
	\label{fig15}
\end{figure}
From Fig.~(\ref{fig15}), it is evident that the condition imposed by Chan et al. \cite{Chan} as given by Eq.~(\ref{eq32}) is well satisfied throughout the interior of the model. 
\subsection{Lagrangian oscillation} To study the stability of our model under small radial oscillation, we graphically represent the variation of the Lagrangian change in radial pressure at the surface of a compact object with frequency $(\omega^{2})$. Pretel \cite{Pretel} introduced the procedure to show the frequency dependence of the Lagrang\\ian perturbation. In this model, we consider $\beta=0$ \cite{Pretel} and the coupled equations demonstrating the radial oscillation are expressed as:\\
\begin{eqnarray}
	\frac{d\zeta}{dr}=-\frac{1}{r}(3\zeta+\frac{\Delta p_{r}}{\Gamma p_{r}})+\frac{d\nu}{dr}\zeta,\label{eq34} \\ 
	\frac{d\Delta p_{r}}{dr}=\zeta\Big(\frac{\omega^2}{c^2}e^{2(\lambda-\nu)}(\rho+p_{r})r-4\frac{dp_{r}}{dr}\nonumber \\-\frac{8\pi G}{c^4}(\rho+p_{r})e^{2\lambda}rp_{r}+r(\rho+p_{r})(\frac{d\nu}{dr})\Big)\nonumber\\-\Delta p_{r}\Big(\frac{d\nu}{dr}+\frac{4\pi G}{c^4}(\rho+p_{r})re^{2\lambda}\}\Big), \label{eq35}
\end{eqnarray} \\ 
where $\zeta$ is the eigen function of the radial part of the Lagrangian displacement and is given by $\zeta=\frac{\delta(r)}{r}$. In this adaptation, $\zeta$ is normalised such that $\zeta(0)=1$. To remove the central singularity in Eq.~(\ref{eq34}), the term with $(\frac{1}{r})$ should vanish as $r\rightarrow0$. Therefore, we obtain the following condition:
\begin{equation}
	\Delta p_{r}=-3\Gamma\zeta p_{r}. \label{eq36}
\end{equation}
At the stellar surface $(r=R)$, the Lagrangian change in radial pressure must also vanish, i.e. 
\begin{equation}
	\Delta p_{r}=0. \label{eq37}
\end{equation} 
\begin{figure}[h!]
	%\centering
	\includegraphics[width=8cm]{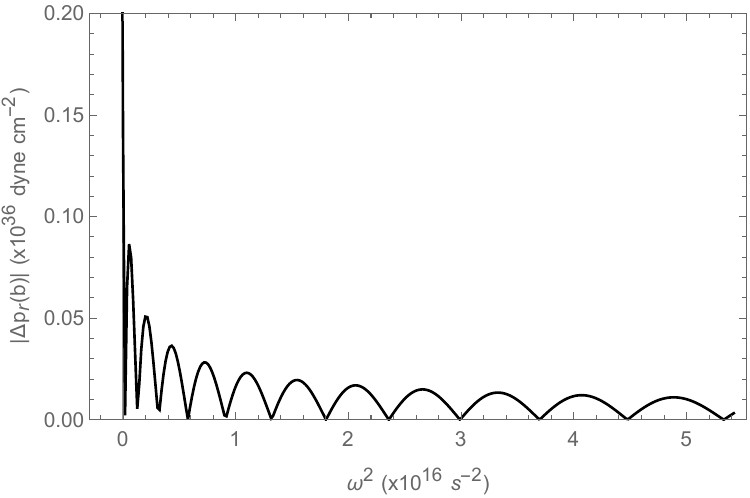}
	\caption{Radial variation of Lagrangian oscillation for $H=0.3$ and $K=0.269\times10^{-7}$.} 
	\label{fig16}
\end{figure}
From Fig.~(\ref{fig16}) it is noted here that $\omega^{2}>0$ for all normal modes of radial oscillation in this model and the correct values of normal frequency modes are characterised by the minima of the plot shown in Fig.~(\ref{fig16}). From the above criteria, we may assert that this model configuration is also stable against small radial oscillatory perturbations.  
\section{Conclusion}\label{sec10}
In this paper, we have presented a new generalised model of anisotropic compact objects by employing the modified Chaplygin equation of state in the context of the Buchdahl-I metric ansatz as described in Ref.~\cite{Delgaty}. For physical analysis, we have studied the basic characteristic properties, i.e., the variations of energy density, radial, tangential pressure and pressure anisotropy,  for the compact object PSR J0740+\\6620 and found that they are suitable for a realistic model. The causality conditions and the energy conditions are well maintained throughout the star. We have also determined the mass-radius relation for compact objects by solving the TOV equation \cite{Tolman,Oppenheimer} along with the MCG EoS, as shown in Fig.~(\ref{fig1}). With the variation of the Chaplygin parameter $H$ as tabulated in Tab.~(\ref{tab1}), when $K=10^{-7}$, the maximum mass of the compact object varies from $1.18 M_{\odot}$ to $3.72 M_{\odot}$. It is evident that the employment of MCG EoS may increase the maximum mass range, which means that compact objects can contain more mass against their gravitational collapse in the presence of MCG. Therefore, it is possible to describe a wide range of mass of compact objects and companion objects of GW events based on recent observations. We have also predicted the radius of the companion star of GW190814, GW170817 events and the pulsars PSR J0952-0607, PSR J2215+5135 and 4U 1608-52, as tabulated in Tab.~(\ref{tab2}). From Tab.~(\ref{tab2}), it is evident that by adjusting the values of $K$ and $H$, one may predict the value of the radius of a compact object of a wide range of mass. However, this model is not suitable for low mass compact objects. This may be explained as follows: the Chaplygin EoS may be considered for exotic matter, which may exist inside a compact object with a higher mass. However, using a suitable combination of $H$ and $K$, one can predict the radius of high mass pulsars and secondary object of GW event 190814. Therefore, as EoS is an important parameter to explain the properties of compact objects, it may be concluded that the interior of such high mass pulsars may contain MCG EoS. To define the stability of the model, we have used the generalised TOV equation, Herrera cracking condition, the adiabatic index variation and radial variation of Lagrangian oscillation. Considering all these arguments, it may be possible to say that our model is indeed a generalised stable and viable representation of an anisotropic compact object.
\begin{acknowledgements}
DB is thankful to the Department of Physics, Cooch Behar Panchanan Barma University, for providing the necessary help to carry out the research work.
\end{acknowledgements}

\section{Declarations}
\begin{description}
\item[Funding:] Not applicable. 
\item[Conflicts of interest:] Not applicable 
\item[Availability of data and material:] This manuscript has no associated data or the data will not be deposited, we have used only obtained mass-radius relations in the context of compact objects to construct relativistic stellar models.
\end{description}

\end{document}